# Making Experiment Results Comparable through Sitewide Impact Estimation

**Authors:** Xiaochen Zhang, Wenjing Zheng, Xingye Kang, Vamsi Manthena

## Abstract

Roblox spans many product surfaces—discovery, the virtual economy, creator tools—that differ sharply in reach and in how much the affected population contributes to any sitewide metric. A randomized experiment shows whether a treatment works for the users in its sample, but this experiment lift is not comparable across surfaces, because surfaces differ in reach and in their contribution to sitewide metrics. The quantity a strategic decision needs is a counterfactual—how much an aggregated sitewide metric (e.g., total platform play time) would move if the treatment reached 100% of its target population, which may be only part of the sitewide population, during the measurement period. We call this sitewide impact. Its scaling factor, which we call the coverage factor, combines the reach and the target population's contribution to the sitewide metric under no treatment. Hence a low-reach surface whose users carry much of the metric could outrank a high-reach surface whose users carry little. This framing is distinct from transportability, which carries an effect to a population with a different covariate distribution: the experiment already measures the effect on the affected population, and we aggregate it over the platform, where users outside the affected population contribute zero and the affected population's share of the metric is measured in the control state. Our target is a relative lift in a sitewide aggregate rather than a per-user average treatment effect, because a per-user effect on the affected population does not account for how many users a treatment reaches.

We develop two estimators under this framing: one for difference-in-means metrics and one for variant-level ratio metrics (a metric defined as a ratio of two population aggregates). Empirically, experiment lift is a poor proxy for sitewide impact—across 139 experiments that are statistically significant on time spent, the Spearman rank correlation between the two is 0.20, and 5 of the 10 largest experiment lifts are among the 10 largest sitewide impacts. We validate the estimator with A/A tests and against a long-term holdout, and apply it to cross-domain prioritization and attributing year-over-year growth to shipped innovations.

## 1. Motivation and Contribution

At Roblox, strategic measurement spans many surfaces—experience discovery, quality of experience, the virtual economy, notifications—each serving different behaviors and objectives. The platform population is correspondingly diverse in geography, device, engagement, context, and role. Experiment surfaces across Roblox may differ greatly in reach, user population, and ability to move certain metrics. Figure 1 shows the distribution of affected sample sizes, baseline metric contribution and treatment responsiveness across eight large product surfaces from 89 live experiments on a snapshot date.

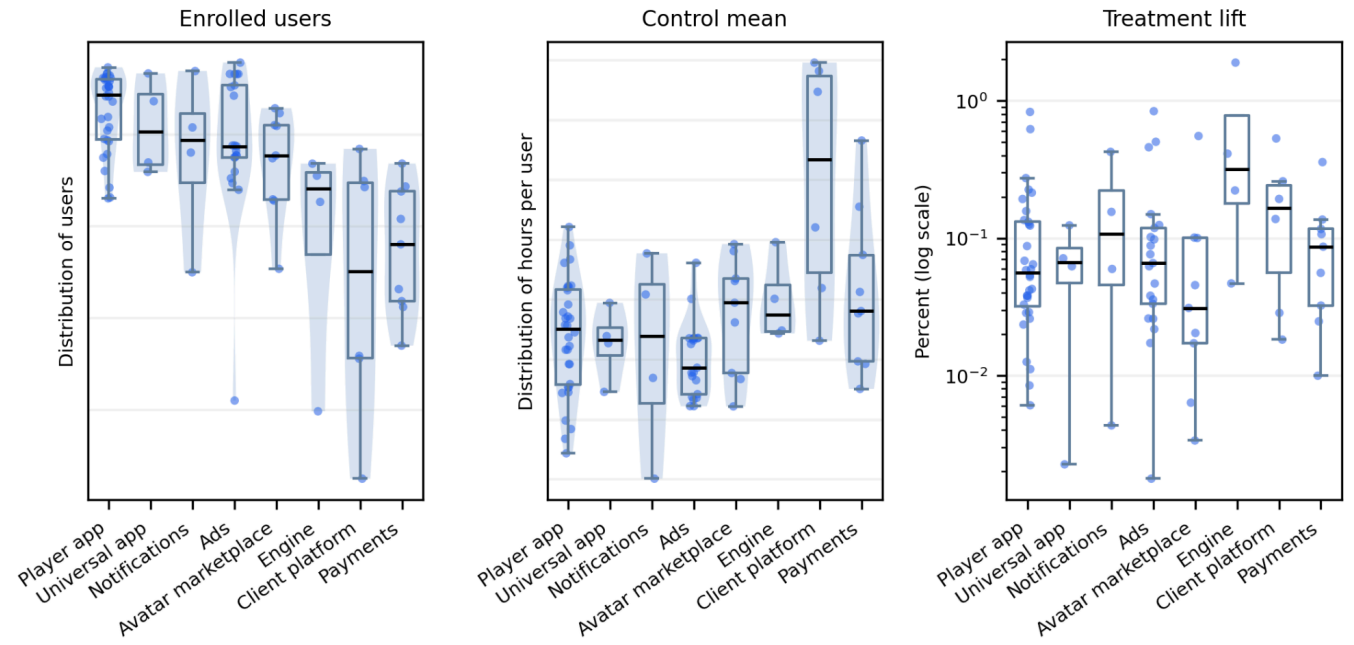

Figure 1. Each point is one A/B experiment that was live on 2026-09-08 and reported a time spent result. Shown are the 89 experiments in eight of the product domains that run the most experiments, out of 148 in total. Boxes give the median and interquartile range across experiments; whiskers extend to 1.5x the interquartile range. Shaded shapes in the first two panels are kernel density estimates of the same experiments; those two panels omit the numeric scale, so only relative spread is read. Domains are ordered by median enrolled users, and all vertical axes are log scaled.

Figure 2 compares experiment lift with sitewide impact across experiments. Even a relatively large experiment lift can produce a much smaller sitewide impact when the affected population's baseline contribution is small. Experiment lift alone cannot compare impact across experiments.

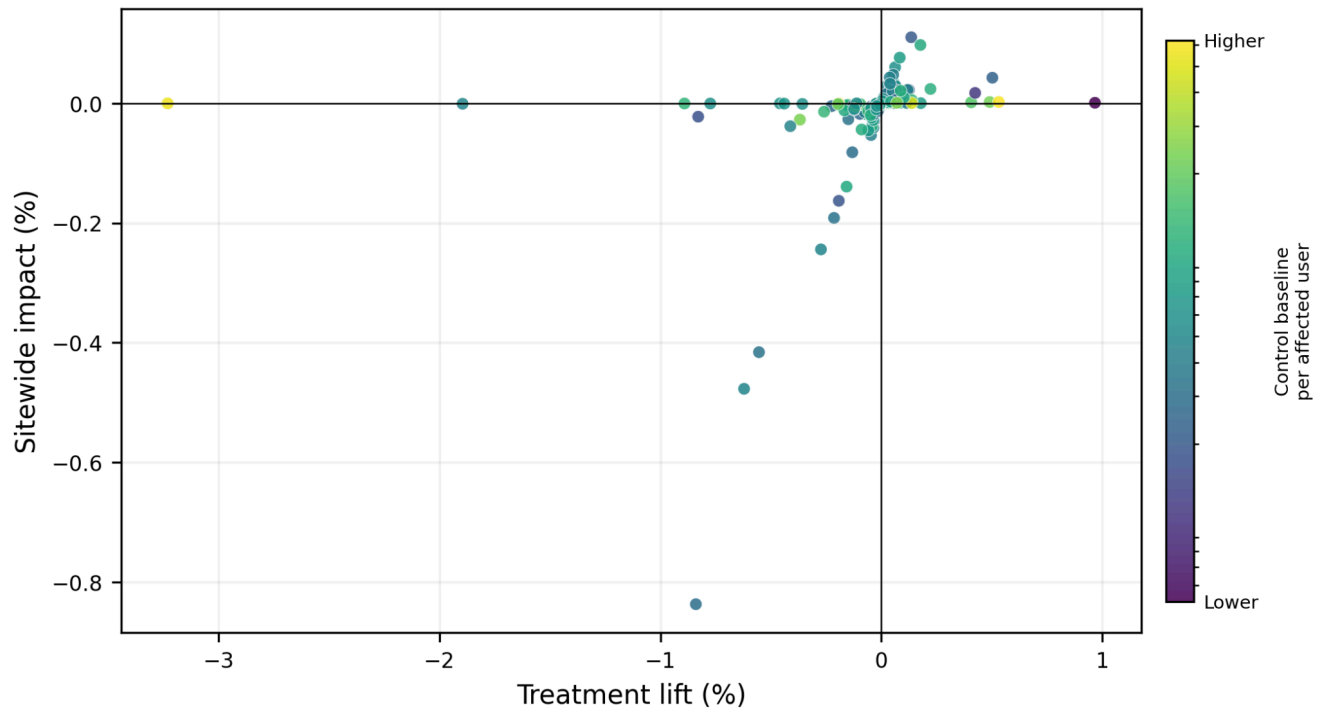

Figure 2 (baseline coloring). The same 148 time-spent experiments as Figure 1. One test (+53% lift, near-zero reach) is omitted. Color is the pre-experiment-adjusted control mean per affected user. The scale is withheld, so color shows relative ordering only, from lower to higher baseline.

Strategic decisions need innovations measured by different experiments to be comparable on the same scale.

This calls for a well-defined counterfactual estimand: how the population's metrics would change if the treatment reached its full target population.

This paper makes four contributions:

- We define sitewide impact as a platform-wide aggregate lift under the counterfactual that the treatment reaches 100% of its target population, measured over a stated sitewide population.
- We derive estimators for difference-in-means and for variant-level ratio metrics as applications of that counterfactual. These estimators adjust for both reach and metric contribution of the experiment surface.
- We extend the construction to heterogeneous effects through a segment-specific estimand.
- We validate the estimator and apply it to

cross-domain ranking and to attributing measured innovations to year-over-year growth.

## 2. Related Work

The closest literature is on transportability and external validity of randomized experiments [1], [2]. In that setting, the study sample and the population of interest differ in their covariate distributions, and the estimand is the average treatment effect in the latter. Identification relies on reweighting or outcome modeling conditional on covariates and requires positivity [9]: every covariate profile in the population of interest must have a nonzero probability of appearing in the trial. Sitewide impact answers a different question. In an online experiment, the users in both arms are a random sample of the target population, which is the population a full rollout would treat. The experiment lift therefore already estimates the effect on the affected population under full rollout, and there is no covariate shift to correct. The question sitewide impact answers is instead how that effect scales into a change in a platform aggregate. Users outside the affected population are not assigned a transported effect: they stay untreated in the counterfactual and contribute zero change. Positivity between them and the affected population fails by construction (in a notification experiment, users who never open the notification center can never enter the affected population), but because no effect is carried over to them, it is not required. Our estimand also differs—a relative lift on a population aggregate rather than an average treatment effect—because a per-user effect cannot rank interventions with different reach. In online experimentation, trigger analysis already rescales the effect on triggered users to the whole population by their share of the metric, which is called the diluted treatment effect [4], [5]. The coverage factor plays the same role. Beyond that, we reconstruct the no-treatment sitewide baseline from experiment data, extend the construction to variant-level ratio metrics, and recombine it across segments.

## 3. Estimand and Sitewide Segments

Let $S$ denote the sitewide population during the measurement period, and let $A \subseteq S$ denote the target population (i.e. users who are reachable and eligible for the treatment). Let Y be the outcome of interest and $Y_i(1)$ and $Y_i(0)$ be user $i$'s potential outcomes under treatment and no treatment. Let $\mu_{S|1}$ and $\mu_{S|0}$ be the counterfactual sitewide means, and $\mu_{A|1}$ and $\mu_{A|0}$ be the counterfactual affected-population mean under the two rollout states.

*3.a) Difference-in-means metric*

For a difference-in-means metric, the sitewide-impact estimand used in production is

$$L_S = \frac{\mu_{S|1} - \mu_{S|0}}{\mu_{S|0}} = \frac{\frac{1}{N_s}\sum_{i\in S}\left[Y_i(1) - Y_i(0)\right]}{\mu_{S|0}} = \frac{\sum_{i\in A}\left[Y_i(1) - Y_i(0)\right]}{N_S\mu_{S|0}}.$$

Let $L_A$ be the affected-population lift.

$$L_A = \frac{\sum_{i\in A}\left[Y_i(1) - Y_i(0)\right]}{\sum_{i\in A} Y_i(0)} = \frac{\sum_{i\in A}\left[Y_i(1) - Y_i(0)\right]}{N_A\mu_{A|0}},$$

where users outside $A$ contribute zero under two assumptions: a) SUTVA [3], and b) reach is unaffected by treatment. On the second assumption: a better product feature may itself draw a higher share of the sitewide users to the feature surface, i.e. A could be different under full productization. Holding reach fixed biases the estimate whenever treatment changes reach (downward in this example), but it gives every product surface the same reference point, the control state, so results stay comparable and easy to interpret.

This yields a relationship between sitewide lift and experiment lift $L_S = L_A C$. And $C = \frac{N_A\mu_{A|0}}{N_S\mu_{S|0}}$ is the coverage factor: the affected population's share of the sitewide metric when no one is treated. For example, a 2% lift in an affected population that holds 10% of the metric is a 0.2% sitewide lift. In the estimand formula, $L_A$, $N_A$, $\mu_{A|0}$, $N_s$ come directly from the experiment. The observed sitewide mean, however, already includes the treatment effect on treated users. To remove that effect and estimate the counterfactual no-treatment sitewide mean $\mu_{S|0}$,

let $\mu_S^{obs}$ be the observed sitewide mean, $N_T$ and $N_C$ the eligible treatment and control counts, and $\bar{Y}_T$ and $\bar{Y}_C$ their outcome means. $\hat{\mu}_{S|0} = \mu_S^{obs} - \frac{N_T}{N_S}(\bar{Y}_T - \bar{Y}_C)$.

*Heterogeneity*

Users are not all alike. Different user segments, such as new users and long-time payers, differ in how much of the segment an experiment reaches, in how much the segment contributes to the metric before treatment, and in how the segment responds to the treatment. Each segment therefore has its own affected population, baseline contribution, and lift, and so its own coverage factor. The overall sitewide lift adds up the segments' sitewide lifts, each weighted by that segment's share of the metric. The same lift matters more sitewide when it lands in a segment with a large coverage factor than in one with a small coverage factor.

For the same reason, averaging segment lifts does not recover the overall lift, because it gives every segment the same weight. Across 155 experiments, combining segment totals instead brings the result much closer to the reported overall lift (Figure 3): the mean absolute error falls from 1.33 to 0.47 percentage points for time spent and from 7.01 to 1.68 percentage points for bookings. The recombined lift is not exact because the supported segments cover a median of 80% of each experiment's users.

To represent heterogeneity, let $g$ index a supported segment, such as new, US, Frequent Android players. The segment-specific estimand is

$$L_{S,g} = \frac{\mu_{S,g|1} - \mu_{S,g|0}}{\mu_{S,g|0}}.$$

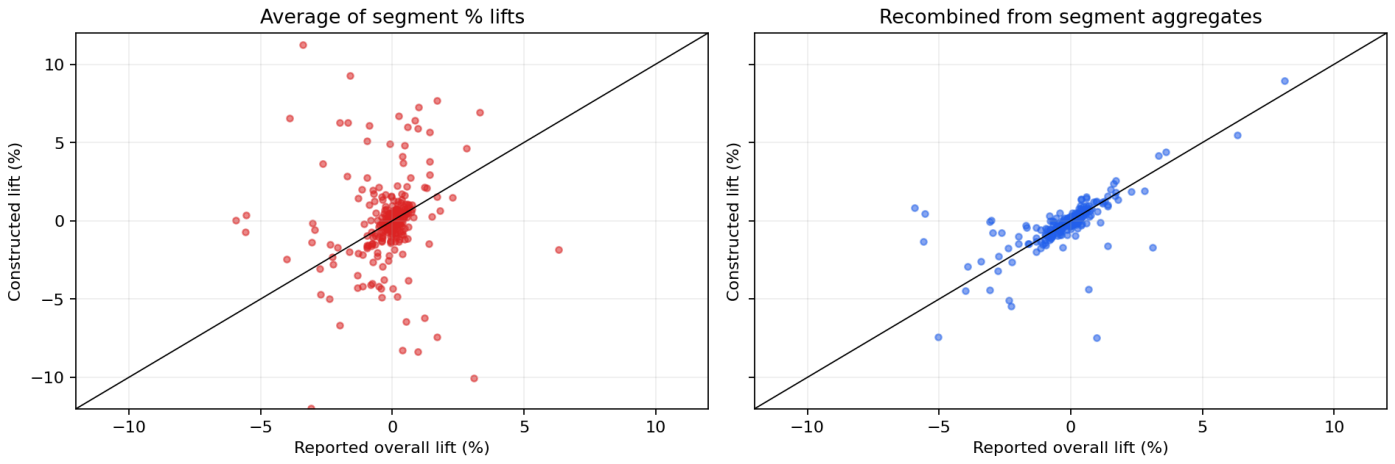

Figure 3. Strategic cohorts on 2025-10-01. The left panel is the unweighted average of the segment percent lifts; the right panel recombines the segment aggregates. Mean absolute error against the reported overall sitewide lift is 1.33 versus 0.47 percentage points for time spent (n=154), 7.01 versus 1.68 for bookings (n=152), and 1.01 versus 0.51 for the purchaser indicator (n=152).

In Figure 3, averaging segment lifts (left) scatters widely around the reported overall lift, while combining segment totals (right) stays close to it. For nonoverlapping[1] segments, the overall lift is recombined from counterfactual aggregate components: $L_S = \frac{\sum_g N_{S,g}\mu_{S,g|1}}{\sum_g N_{S,g}\mu_{S,g|0}} - 1.$

*3.b) Variant-level ratio metric*

Many platform metrics are ratio metrics [6], such as crashes per hour of play. A ratio metric can be defined at two levels. A user-level ratio metric computes the ratio for each user and then averages it across users, so every user counts equally. A variant-level ratio metric divides the total numerator by the total denominator over all users in a variant, so users who play more count more. The two can differ widely. Take one user who has one crash in one hour of play and another who has no crashes in nine hours. At the user level they average 0.5 crashes per hour; at the variant level the rate is 0.1 crashes per hour. The variant-level ratio metric is the one that matches platform totals: total crashes divided by total play time is the platform's crash rate. That is the one we derive.

A user-level ratio metric is a per-user average, so the coverage factor from Section 3.a applies to it directly. A variant-level ratio metric is not a per-user average, and neither the numerator's coverage factor nor the denominator's gives its sitewide lift. The affected population holds a different share of crashes than of play time. Scaling the experiment lift of the ratio metric by either share gives a different answer, and neither equals the change in total crashes over total play time. Across 484 experiments that ended between September 8 and October 29, 2026 (crash metrics are available from September 8), the two single-factor answers differ from each other by a median of 3.1% of the sitewide lift. Instead, we apply each coverage factor to its own total and then take the ratio.

For a variant-level ratio metric, let $Num$ and $Den$ be the numerator and denominator components. Reconstruct the no-treatment components as

$$\hat{Num}_{S|0} = Num_S^{obs} - \frac{N_T}{N_S}(\overline{Num_T} - \overline{Num_C})$$

$$\hat{Den}_{S|0} = Den_S^{obs} - \frac{N_T}{N_S}(\overline{Den_T} - \overline{Den_C}).$$

Estimate the affected-population component changes directly:

$$\Delta\hat{Num}_S = \frac{\hat{N}_A}{N_S}(\overline{Num_T} - \overline{Num_C})$$

$$\Delta\hat{Den}_S = \frac{\hat{N}_A}{N_S}(\overline{Den_T} - \overline{Den_C}).$$

The sitewide lifts for ratio metrics are

$$\hat{M}_{S|0} = \frac{\hat{Num}_{S|0}}{\hat{Den}_{S|0}}$$

$$\hat{M}_{S|1} = \frac{\hat{Num}_{S|0} + \Delta\hat{Num}_S}{\hat{Den}_{S|0} + \Delta\hat{Den}_S}$$

$$\hat{L}_S^{ratio} = \frac{\hat{M}_{S|1}}{\hat{M}_{S|0}} - 1.$$

In short, we remove the treatment effect from the observed numerator and denominator to get their no-treatment values, add the effect as if the whole affected population were treated, and compare the ratio after with the ratio before. Figure 4 shows what goes wrong if one coverage factor is used for the whole ratio instead. With the denominator's factor, the median error is 3.5% of the sitewide lift, and 29% or more for one experiment in ten. With the numerator's factor, the error is 7.2% or more for one experiment in ten.

*3.c) Intervals for sitewide lift*

Sitewide lift is experiment lift times the coverage factor C, so an interval for experiment lift carries over to sitewide lift by the same scaling. The platform reports a 95% interval for the difference in means. We divide both ends by the control mean to express it as a relative lift, and then multiply by C. This treats both the control mean and C as known. C is a ratio of totals over the whole sitewide population, so its sampling error should be small next to that of the experiment lift, but we have not measured it, and the interval is slightly too narrow as a result.

When a result is read every day, as with the holdout in Section 4.c, repeated looks make a fixed 95% interval too narrow. We widen it with the Lan–DeMets approximation to an O'Brien–Fleming boundary [7], [8]: on day t the critical value is $z_t = \frac{1.96}{\sqrt{n_t/n_{end}}}$, where $n_t$ is the sample size on day t and $n_{end}$ the sample size on the last day. The interval is wide in the first days and returns to ±1.96 standard errors on the last day. Because the last-day sample size is known only once the holdout ends, this is a band for reading the result afterwards, not a stopping rule.

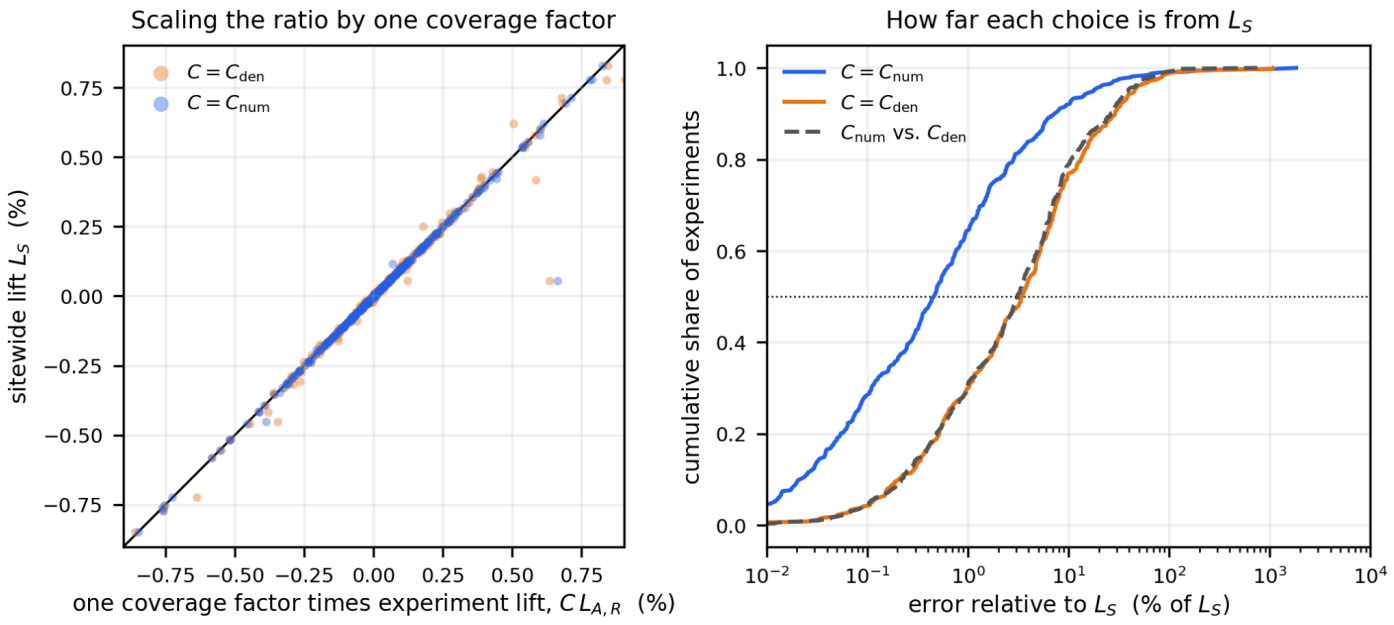

Figure 4. Crashes per hour of play on 484 Classic A/B experiments that ended between September 8 and October 29, 2026, with crashes as the numerator and play time as the denominator. Each point scales the experiment lift of the ratio metric by one coverage factor. With $C_{\mathrm{den}}$ the median error is 3.5% of $L_S$, and 29% or more for one experiment in ten; with $C_{\mathrm{num}}$ it is 7.2% or more for one in ten. The two choices differ from each other by a median of 3.1% of $L_S$.

## 4. Empirical Results and Strategic Applications

### *4.a) Finding: experiment lift does not track sitewide impact*

The estimator is integrated into Roblox's internal experimentation platform. We find that ranking by experiment lift does not recover ranking by sitewide impact, which is why $\hat{L}_s$ is needed for prioritization.

Among the 139 A/B tests that started and ended in 2026 and are statistically significant on time spent (two-sided $p<0.05$), Spearman's rank correlation between $|\hat{L}_s|$ and $|\hat{L}_A|$ is 0.20. Figure 5 shows that of the 10 largest experiment lifts, 5 are also among the 10 largest sitewide impacts (50%, versus 10/139≈7% under independent rankings). For top-k lists with k up to 50, 35% to 50% of the experiments in one list are also in the other. Experiment lift is therefore not a reliable proxy for which experiments matter sitewide.

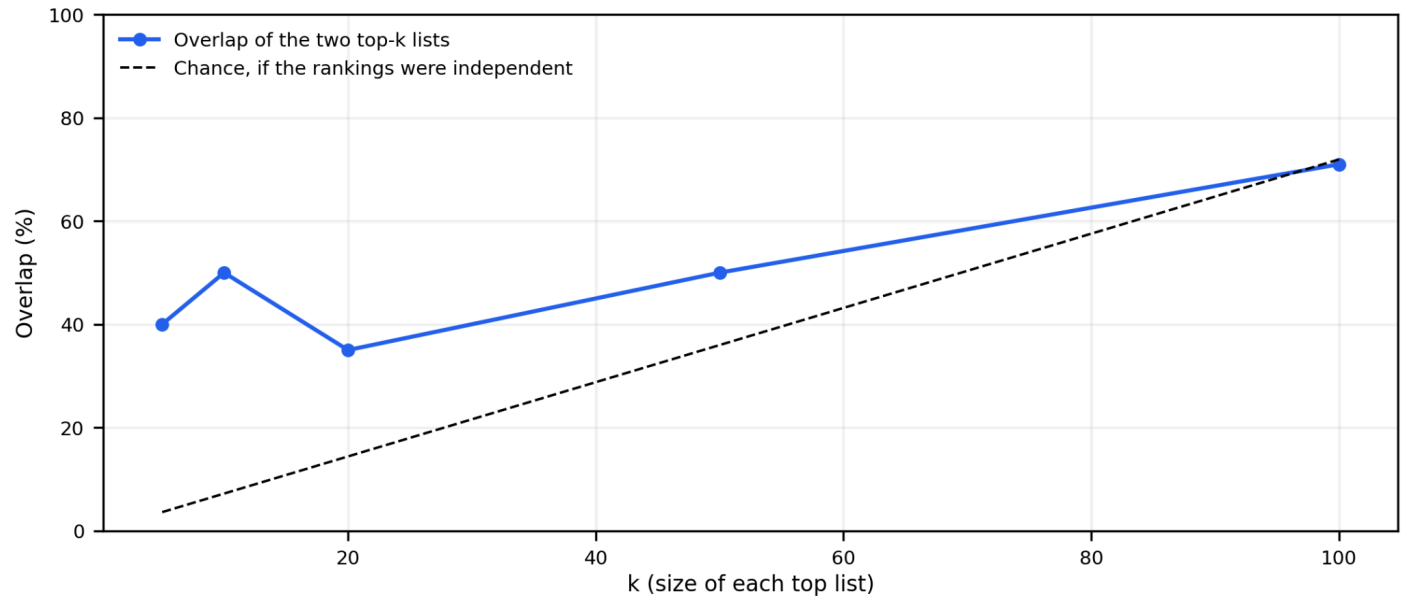

Figure 5. Ranking by |treatment lift| and by |sitewide impact| on 139 Classic AB experiments that started and ended in 2026 and are statistically significant on time spent (two-sided unadjusted p<0.05). Each point is the share of the local top-k that is also in the sitewide top-k.

### *4.b) Validation with A/A tests*

In an A/A test, both groups get the same experience, so the true lift is zero. We use the pre-experiment period of 1,349 experiments that closed in 2026 as A/A tests on time spent, after dropping 64 whose coverage factor exceeds 1, which is impossible for a share and marks a data error, and 27 with fewer than 10,000 pre-period users in an arm. The median sitewide lift is essentially zero (0.00004%), and a signed-rank test does not reject zero ($p = 0.59$, Figure 6), as expected if the estimator adds no bias. Its spread is much smaller than that of experiment lift, but this follows from the scaling: the median coverage factor is 34%, which shrinks noise and effect alike, so the narrower spread is not extra precision.

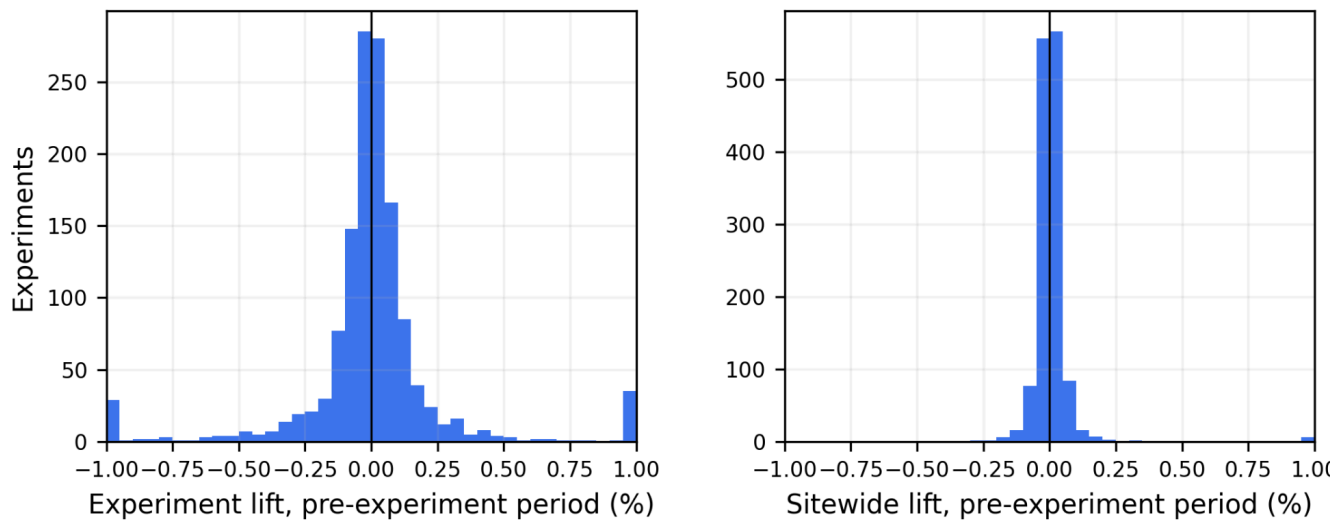

Figure 6. A/A test that uses the pre-experiment period as the metric, for 1349 closed 2026 A/B experiments (time spent). Sitewide lift is experiment lift × coverage factor. The median coverage factor is 34%, so the sitewide distribution is narrower by construction. Bars are clipped at ±1%; 60 experiments fall outside that window. Excluded: 64 with a coverage factor above 1 (data errors) and 27 with fewer than 10,000 pre-period users in an arm.

### *4.c) Validation against a long-term holdout*

A holdout test withholds a change from a random share of the population, so it measures the 100%-rollout counterfactual directly. We use one to validate $\hat{L}_s$ : the sitewide impacts of the experiments a holdout covers should sum to the lift the holdout reports over the same window.

Our case is a payments holdout that ran from January 20 to March 18, 2026 across six payment surfaces. Eleven A/B tests started and finished on those surfaces in that window; keeping only the latest version of three experiments that were rerun leaves eight innovations, at least one per surface. Summing their impacts assumes no interference across surfaces (SUTVA [3]). Figure 7 shows bookings (winsorized at the 99.9th percentile) and the probability of spending Robux. The blue line is the holdout-implied lift, and the band is its 95% interval, built as described in Section 3.c. The red step line adds each innovation's sitewide impact at its end date, using, for each metric, the candidate arm with the highest sitewide lift, because the platform does not record which arm shipped. The red line is therefore an upper bound. The shaded red area runs from the sum of the worst arms to the sum of the best. It widened in March because two experiments, the ML redirect variants and the recommendation model, had arms with clearly negative sitewide lift; the holdout ends at or above the top of the range on both metrics, consistent with the better arms being the ones that shipped. The stacked series runs ahead of the holdout through February, likely because each impact is counted on its experiment's end date before the change is fully rolled out, and ends inside the holdout's interval on both metrics: 0.41% against a holdout lift of 0.50% for bookings, and 0.22% against 0.21% for the probability of spending Robux. This supports the aggregate magnitude of $\hat{L}_s$ over this window.

We also checked two engagement metrics, time spent and DAU (Figure 8). The holdout shows a small negative sitewide lift on both, −0.025% for time spent and −0.003% for DAU, while the stacked series is higher: across candidate arms it ranges from about 0% to +0.03% for time spent and from −0.001% to +0.010% for DAU. The holdout's 95% interval is [−0.070%, +0.021%] for time spent and [−0.022%, +0.017%] for DAU. Both include zero, so the negative holdout values are within noise. For DAU the whole stacked range lies inside the interval; for time spent the low end does, while the best-arm value of +0.029% sits just above it. The platform's guardrail tests, which check for a drop larger than 0.28% in time spent and 0.12% in DAU among holdout users, pass on both. These payments5 changes were not expected to move engagement, and the holdout is consistent with that.

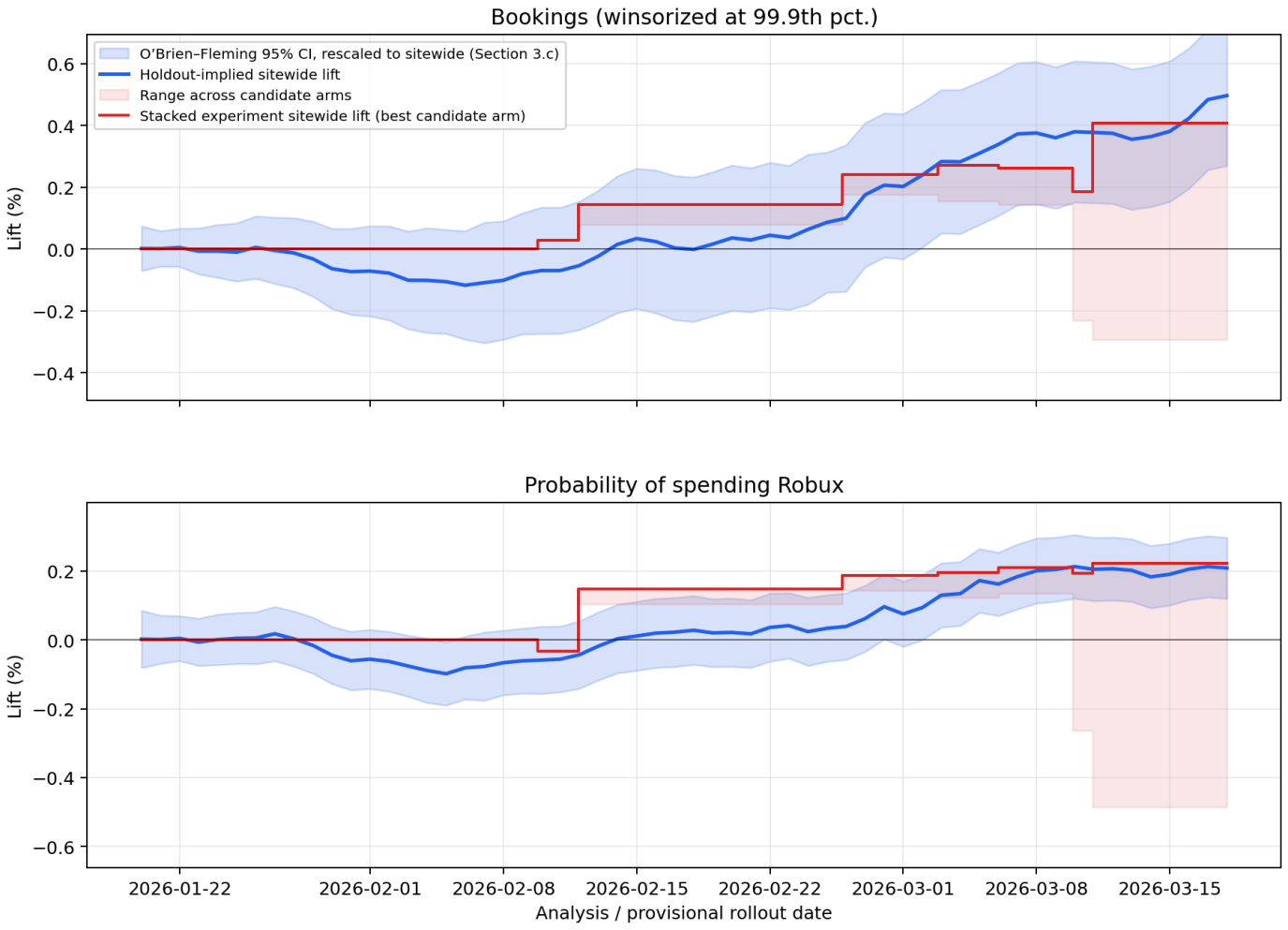

Figure 7. Stacked experiment sitewide impact versus holdout-implied lift.
The blue band is the holdout's O'Brien-Fleming 95% interval, rescaled by coverage and built as described in Section 3.c.

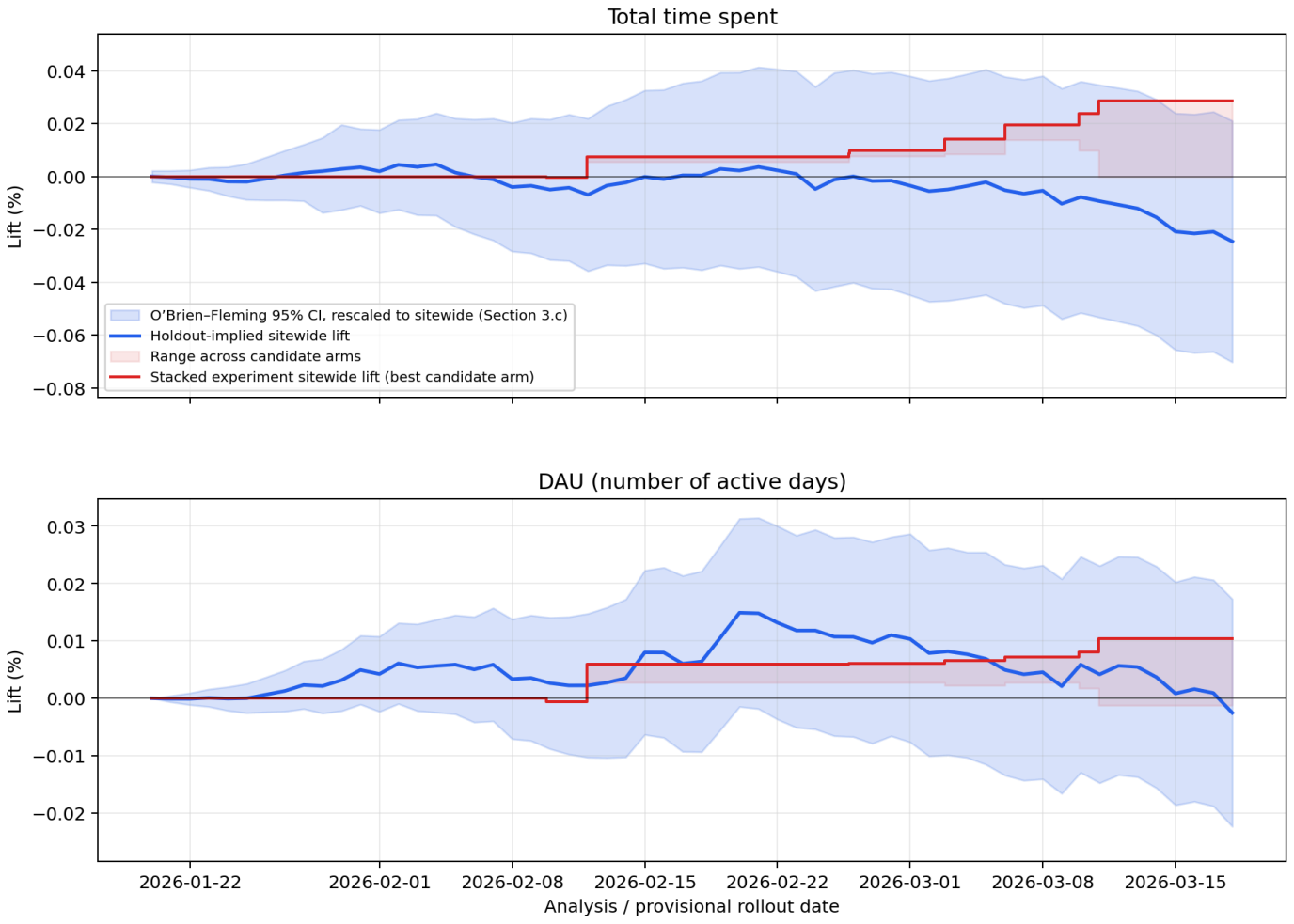

Figure 8. Engagement metrics: stacked experiment sitewide impact versus holdout-implied lift.
The blue band is the holdout's O'Brien-Fleming 95% interval, rescaled by coverage and built as described in Section 3.c.

### *4.d) Strategic applications*

Sitewide impact supports two strategic uses. The first is cross-domain prioritization: because sitewide lift expresses every intervention as a change in the same platform aggregate under the same counterfactual, experiments on different surfaces can be ranked against one another on a given metric. The second is growth attribution. An internal method built on sitewide lift decomposes observed year-over-year growth into organic growth and a contribution from product launches. It takes the sitewide lifts of long-term holdouts with non-overlapping feature sets, estimates organic growth from the holdouts' control groups, and computes the product contribution as the sum of the holdout lifts scaled by one plus the organic growth rate. The decomposition assumes that interactions among launches are negligible and that each holdout's features are rolled out to the full population when the holdout ends.

## 5. Limitations and Conclusion

The estimand rests on two strong assumptions that we restate here. First, SUTVA: a user's outcome depends only on their own assignment; it can fail when a treatment given to one user changes the outcomes of other users. Second, reach is unaffected by treatment: the affected population is treated as fixed, whereas a strong treatment may itself change who is in it. Both bias the sitewide estimate when violated, and neither is testable from the experiment alone.

Computation is also costly at scale: the counterfactual baseline is experiment-specific (all sitewide populations are re-aggregated for each experiment window), and difference-in-means and ratio metrics need separate estimators. We precompute fixed lookback windows, which fit long-running experiments poorly. Relaxing the assumption that reach is unaffected by treatment, handling interference, and improving scalability are left to future work.

In summary, sitewide impact connects local causal evidence to platform-wide decisions. A single counterfactual, applied separately to difference-in-means and ratio metrics and extended to segments, yields estimates that are comparable across surfaces and usable for prioritization and growth attribution.